\documentclass[12pt]{iopart}
\usepackage{dcolumn}
\usepackage{graphicx}

\usepackage{breqn}

\newcommand{\ii}{\textbf{i}}
\newcommand{\ee}{\textbf{e}}

\graphicspath{{./} {figures/}}

\begin{document}

%opening
\title[iSURF: infinite-time surface flux methods]{iSURF: A family of 
infinite-time surface flux methods}
% TODO 

\author{F. Morales$^1$, T. Bredtmann, and S. Patchkovskii$^1$}
\address{$^1$Max-Born-Institute, Max-Born-Strasse 2A, D-12489 Berlin, Germany}
\ead{serguei.patchkovskii@mbi-berlin.de}
\vspace{10pt}

\begin{abstract}
The computation and analysis of photoelectron spectra (PES) is a fundamental technique in atomic and molecular physics to study the structural and dynamical properties of a target system, and to gain insight into the process of its ionization. Since the first numerical solutions of the time-dependent Schr\"odinger equation, numerous methods have been developed to extract PES from the calculated wave functions. However, most of these methods have severe limitations or are computationally very demanding. Here we present a new family of methods, based on the ideas of the so-called analytical Volkov continuation, or time-dependent surface flux (\cite{Ermolaev99a,Ermolaev00,tao2012}), that allows one to obtain fully-converged PES at the end of the laser pulse using either Volkov states or the exact scattering-states, and that has been implemented in the Time Dependent Schr\"odinger Equation (TDSE) solver \cite{Patchkovskii2016}.
\end{abstract}

\section{Introduction}

%The development of ever more sophisticated techniques for detecting energy
%and angle resolved photoelectron spectra (PES) allows for an increasingly detailed insight 
%not only into the electron structural and dynamical properties of the target system, 
%but also into the process that is associated with the ionization step.

Angle- and energy-resolved photoelectron spectroscopy is among the most accurate and popular techniques in atomic and molecular science \cite{wu2011}.
The most basic method to analyze photo-ionized electrons is the time-of-flight spectroscopy, providing a
high resolution in energy of the electron. State-of-the-art technologies, such as 
VMI (Velocity Map Imaging, \cite{chandler87,bordas96}) and COLTRIMS (Cold Target Recoil Ion 
Momentum Spectroscopy \cite{dorner00}) yield both energy and angular resolved PES.
%with resolutions never achieved before.
These experimental techniques have been successfully applied to study processes involving ionization, 
e.g., above threshold ionization (ATI)~\cite{agostini79} or sequential and non-sequential double ionization 
of atoms~\cite{weber00,moshammer00,feuerstein01}, and facilitate advanced experimental methods 
such as time-resolved photoelectron holography~\cite{huismans11}, or the attoclock setup~\cite{eckle08,torlina2015}.
In addition, photoelectron spectra underlay characterization of ultrashort laser pulses 
or pulse trains, using techniques such as RABITT (reconstruction of attosecond beating by 
interference of two-photon transition)~\cite{paul01} and the attosecond streak camera~\cite{itatani02}.

As the complexity and resolution of experimental PES has
increased, calculation of highly accurate PES has become essential, triggering
advances in theoretical methods both for solving the time-dependent
Schr\"odinger equation ( see for example \cite{Tong97a,Nirhuda99a,Muller99a,Borisov01a,Bauer05a,Guan09a,Sorevik09a,Tong11a,Dziubak12a,Shen13a,Broin14a,Patchkovskii2016}) and extracting accurate
photoelectron spectra.  The commonly accepted formal definition of the PES for
1-electron ionization is the projection of the solution of the TDSE onto
asymptotic scattering states of the binding potential in the absence of the
perturbation.  Although asymptotic scattering states are known analytically for
Coulomb potentials, numerically stable calculation of these states is not
entirely trivial \cite{barnett1982}. For a more general binding potential,
the calculation of a large number of scattering states and the corresponding
projections may become quite cumbersome, spurring the development of numerous
alternative approaches to calculate PES .

Among the most popular techniques are:
\begin{itemize}
\item Window methods \cite{schafer90}: This method uses a energy-dependent 
window function to project onto the real field-free Hamiltonian of the system.
Here, only the absolute value of the PES amplitude is retrieved. Due to the 
use of the window function, the spectral resolution is limited, especially in
the low-energy region where high Rydberg states need to be separated from the
true continuum. The wavefunction needs to remain inside the simulation volume 
until the end of the ionizing laser pulse. 
Among the major advantages of the window methods is the correct handling of long-range
potentials and norm conservation.

\item Coordinate space masking + Fourier Image: This commonly used method is 
equivalent to the projection onto a plane-wave continuum.  It requires that the
continuum and bound parts of the wavefunction are well-separated in coordinate
space.  Similar to the window methods, the wavefunction must remain within the
simulation volume, in which case the technique is norm-conserving.  For
long-range potentials, the photoelectron energies are overestimated.  Spurious
interferences may also arise due to the mixing of contributions from different
energies to the same plane wave $k$ vector. Finally, artifacts due to the masking
step are difficult to avoid.

\item Numerical calculation of scattering states (see \cite{Palacios07,Feshchenko13,Masataka15} for recent examples): 
This is an exact method that provides exact spectra, with the 
correct scattering phase, as long as the correct scattering states are known. 
For arbitrary potentials, calculating a numerical solution of these 
functions is a delicate task. 
However, this method is not norm conserving for any finite $k$ grid.
The wavefunction needs to remain within the simulation volume.

\item Volkov-state continuation (also known as tSURFF)\cite{Ermolaev99a,Ermolaev00,Serov01a,tao2012}: 
This method is designed to remove the requisite of retaining the entire wavefunction
remaining within the simulation volume, which might become prohibitively expensive for the
case of an intense infrared laser field.  In this approach, the ponderomotive potential
and the free-electron oscillation amplitude are assumed to be large.  Here,
 the projection onto a scattering state is replaced by a time integral
of the outer-surface flux, thus allowing much smaller simulation volumes.  In
common with the scattering-state projection tSURFF is not norm-conserving.  In
common with the Fourier image approach, it is only rigorous for a short-range
potential, and introduces similar artifacts in the PES (see below). Calculation 
of converged PES may also require long field-free propagation after the end of the pulse,
especially if the low-energy parts of the spectra are desired \cite{Zielinski16} (see below).

\end{itemize}

In order to eliminate boundary reflections, most of the practical approaches to solve the TDSE include the use of
a Complex Absorbing Potential (CAP) (\cite{Riss93, Riss96, Rescigno97,Moiseyev98, Manolopoulos02, Poirier03, Sajeev05, Scrinzi10}), making the Hamiltonian non-Hermitian.
This non-Hermiticity is often viewed as a necessary evil to
keep the computation tractable. Here, we will demonstrate a new
family of methods, complementary to the ideas of \cite{Ermolaev99a,Ermolaev00} and
\cite{tao2012}, which uses the non-Hermiticity of the field-free Hamiltonian to
analytically extend the tSURFF time integral to an infinite time. Used with the
Volkov states, our approach allows calculation of fully-converged tSURFF spectra 
immediately at the end of the laser pulse (from here on referred as the iSURFV method). 
Apart from the time-integral convergence, this technique shares
the advantages and shortcomings of the Volkov-state continuation approaches described above.

For simulation volumes large enough to contain the entire wavefunction at the
end of the laser pulse, the non-Hermiticity of the field-free Hamiltonian
also allows calculation of the exact scattering-state projections. The knowledge
of the asymptotic form of the target state is sufficient for the
calculation of the projection. This method we refer to as the iSURFC method.

The rest of this paper is structured as follows: we start by restating the main expressions of the
tSURFF approach. Then we introduce the analytical continuation to infinite 
time, and demonstrate its results in a series of examples, 
comparing the two implemented target functions (Volkov and Coulomb states) with a naive implementation 
of the original tSURFF method.
  
\section{Theory}

We begin by recapitulating the derivation of the surface-flux approach to 
calculate PES.
The technique was originally developed by Ermolaev et al. \cite{Ermolaev99a,Ermolaev00} and 
Serov et al. \cite{Serov01a},
and popularized by Tao et al. \cite{tao2012}. The discussion is this section closely follows Ref.\cite{tao2012}.

We assume that the exact solutions $\Psi\left(t\right)$ for the 
time-dependent Hamiltonian $\hat{H}$ are known, and satisfy the 
time-dependent Schr\"odinger equation (TDSE):
\begin{equation}
 i \hbar\frac{\partial}{\partial t}\Psi = \hat{H}\Psi. \label{eqn:href}
\end{equation}
We are interested in expressing $\Psi$ in terms of the solutions 
$X\left(t\right)$ of a (possibly different) time-dependent Hamiltonian 
$\hat{H}_{S}$.
Functions $X$ satisfy the corresponding TDSE:
\begin{equation}
 i \hbar\frac{\partial}{\partial t}X = \hat{H_S}X, \label{eqn:hsbar}
\end{equation}
or, equivalently:
\begin{equation}
 - i \hbar\frac{\partial}{\partial t}X^* = X^*\hat{H_S}. \label{eqn:hsbar:left}
\end{equation}

The projection of $\Psi$ onto $X$ is given by:
\begin{equation}
 a_X(t_0) = \int d\vec{r} \{ X^* (\vec{r},t_0) \Psi_S (\vec{r},t_0) \} 
\label{eqn:projection}
\end{equation}
For example, if $X$ belongs the continuum part of the spectrum, $a_X$ is the 
corresponding ionization amplitude.

For sufficiently large $t_0$, the projection (\ref{eqn:projection}) onto a 
continuum state 
is dominated by the part of space far away from the coordinate origin, and can be 
replaced by:
\begin{equation}
a(t_0) =_{t_0 \longrightarrow\infty} \int d\vec{r} \{ X^* (\vec{r},t_0) 
\hat{\Theta_S}\Psi (\vec{r},t_0) \}, \label{eqn:projection:gated}
\end{equation}
with $\hat{\Theta}_S$ being the spherical Heavyside function of r:
\begin{equation}
\hat{\Theta}_S=\cases{0,& $r < R_0$\\1,& $r \geq R_0$\\}. \label{eqn:heavyside}
\end{equation}
If we further assume that at some initial time $t_i$ the wavepacket $\Psi$ is 
localized near the origin, so that:
\begin{equation}
 \hat{\Theta}_S \Psi(\vec{r},t_i) \equiv 0 \label{eqn:localized}
\end{equation}
eq.~\ref{eqn:projection:gated} can be re-written as a time integral:
\begin{eqnarray}
 a(t_0) & = \int^{t_0}_{t_i} dt \frac{\partial}{\partial t} \int dr 
X^*(\vec{r},t) \hat{\Theta}_S \Psi(\vec{r},t) \nonumber \\
        & = \int^{t_0}_{t_i} dt \frac{\partial}{\partial t} \int dr 
\frac{i}{\hbar} X^* \{ \hat{H}_S \hat{\Theta}_S - \hat{\Theta}_S \hat{H} \} \Psi.
     \label{eqn:projection:time}
\end{eqnarray}

Provided that the Hamiltonians $\hat{H}$ and $\hat{H}_S$ coincide outside of the 
central region ($r\geq R_0$), the term in the curly brackets in 
eq.~\ref{eqn:projection:time} becomes a commutator 
$\left[\hat{H}_S,\hat{\Theta}_S\right]\equiv\left[\hat{H},\hat{\Theta}_S\right]
$. This
commutator is non-zero on the surface of the dividing sphere, and vanishes 
identically everywhere else.
For the specific case of the dipole-approximation Volkov Hamiltonian in velocity 
gauge:
\begin{eqnarray}
  \hat{H}_S = \frac{1}{2m} (\hat{p} - e\vec{A})^2 = \frac{1}{2m} (-i \hbar 
\hat{\nabla} - e\vec{A})^2, \label{eqn:volkov-hamiltonian}
\end{eqnarray}
the volume integral in eq.~\ref{eqn:projection:time} becomes:
\begin{eqnarray}
 b\left(t\right) & = \int dr X^* \frac{i}{\hbar} [ \hat{H}_S,\Theta] \Psi 
\nonumber \\
                 & = \int\limits_{R_0} d\Omega_r r^2 \{ \frac{i\hbar}{2m} 
(\frac{\partial X^*}{\partial r} \Psi  
                           - X^*\frac{\partial \Psi}{\partial r} ) - \frac{e}{m} 
A_r X^* \Psi \},
    \label{eqn:projection:surface}
\end{eqnarray}
where $A_r\left(t\right)$ is the spherical radial component of the vector potential of 
the laser field $\vec{A}$. 
Note that this derivation does not require the target functions $X$ 
to be defined in the
same Hilbert space as the wavepacket $\Psi$. This property makes it possible to 
directly calculate ionization
amplitudes using an $L^2$ representation of $\Psi$.

We would like to emphasize that up to this point, we have recapitulated the formalism described 
in \cite{Ermolaev99a,tao2012}
(see Ref.~\cite{Ermolaev00} for an extension to the eikonal-Volkov Hamiltonian).

\subsection{Analytical continuation of the time integral}

Calculation of ionization amplitudes using 
eqs.~\ref{eqn:projection:time}--\ref{eqn:projection:surface} requires that
the \textit{entire} continuum part of the wavefunction passes through the 
dividing surface $S$. This may require that
the simulation continues long after the pulse is over, especially if low final 
momenta are of interest.
However, any wavepacket can be trivially expanded over the eigenfunctions of the 
field-free Hamiltonian,
giving its evolution at all future times analytically.
Using the same non-Hermitian spherical-coordinate representation as in 
Ref.~\cite{Patchkovskii2016}:
\begin{eqnarray}
  \Psi(t) & = \frac{1}{r} \sum_{LMj} f_{LMj} \Phi^{R}_{LMj}\left(r\right) 
Y_{LM}\left(\Omega_r\right)
             \exp\left(-\frac{\ii}{\hbar}\mathcal{E}_{LMj} (t-t_x)\right),
   \label{eqn:wavefunction:analytic} \\
  \mathcal{E}_{LMj} & = E_{LMj} - \frac{\ii}{2}\Gamma_{LMj}
    \label{eqn:complex:energy}, \\
  f_{LMj} & = \int d r \int d\Omega_r Y_{LM}^*\left(\Omega_r\right) 
\Phi^{L}_{LMj}\left(r\right) \Psi(t_x),
    \label{eqn:eigenstate:amplitude}
\end{eqnarray}
where $\Phi^{R}_{LMj}$ and $\Phi^{L}_{LMj}$ are respectively the right and left 
eigenvector of the field-free
Hamiltonian, associated with the complex energy $\mathcal{E}_{LMj}$. $E_{LMj}$ 
and $\Gamma_{LMj}$ 
are the energy and the lifetime of the state $LMj$. Finally, $L$ and $M$ are the 
usual orbital quantum numbers
and $j$ is the ordinal number of the state within each $L,M$ channel. 
The finite lifetimes $\Gamma_{LMj}>0$ are the consequence of the 
population being absorbed by the
boundary. Eq.~\ref{eqn:wavefunction:analytic} is applicable for all $t\geq 
t_x$. 

Substituting eq.~\ref{eqn:wavefunction:analytic} into 
eqs.~\ref{eqn:projection:time}--\ref{eqn:projection:surface},
we readily obtain:
\begin{eqnarray}
 a\left(\infty\right) - a\left(t_x\right) & = \int_{t_x}^{\infty} \sum_{LMj} 
g_{LMj} 
    \exp\left( \frac{\ii}{\hbar} \left(\epsilon - \mathcal{E}_{LMj}\right) 
(t-t_x) \right) dt,
  \label{eqn:analytical:amplitude} \\
 g_{LMj} & = \int dr X^* (r,t_x) \frac{i}{\hbar} [ \hat{H}_S,\Theta ] 
\frac{1}{r} f_{LMj} \Phi^{R}_{LMj} (r) Y_{LM} (\Omega_r),
  \label{eqn:eigenstate:factor}
\end{eqnarray}
where $\epsilon$ is the energy of the target state $X$ and the quantities $g_{LMj}$ 
are time-independent.
The time integral in eq.~\ref{eqn:analytical:amplitude} converges, provided that 
$\Gamma_{LMj}>0$
for all $L,M,j$ (state in the continuum) or the corresponding amplitude 
$f_{LMj}$ vanishes (bound state),
yielding the final expression:
\begin{equation}
 a\left(\infty\right) - a\left(t_x\right) = \ii \sum_{LMj} \frac{\hbar 
g_{LMj}}{(\epsilon - E_{LMj}) + \frac{\ii}{2}\Gamma_{LMj}}.
  \label{eqn:projection:infinity}
\end{equation}
Note that eq.~\ref{eqn:projection:infinity} defines the overall amplitude as a 
coherent sum of Lorentzian line profiles,
each associated with an eigenstate of the non-Hermitian field-free Hamiltonian 
of the system.

So far, eq.~\ref{eqn:projection:infinity} does not assume any particular form of 
the long-range Hamiltonian $\hat{H}_S$ or
the associated functions $X$. If we choose $\hat{H}_S$ to be the Volkov 
Hamiltonian, implying that the functions $X$ are
plane waves, we obtain an infinite-time correction to the surface-flux integral 
(eq.~\ref{eqn:projection:time}) of Refs.~\cite{Ermolaev99a,tao2012}.
This is the ``iSURFV'' method.

On the other hand, if $\hat{H}_S$ is given by the Coulomb Hamiltonian and thus $X$ are the 
Coulomb scattering functions, eq.~\ref{eqn:projection:infinity}
yields \textit{exact} ionization amplitudes in the presence of a long-range 
potential. This is the ``iSURFC'' method.

The following section gives some examples, illustrating applications of both 
techniques. The technical details of the
implementation in the \textsc{SCID-TDSE} code (\cite{Patchkovskii2016}) given in Appendices A and B.

\section{Illustration examples}
\subsection{Few-photon ionization of the Hydrogen atom ground 
state\label{sec:example:hfew}}

We consider the hydrogen atom in its $1s$ ground state. The laser field is a 
linearly-polarized Gaussian pulse along the Cartesian $Z$
direction, with a peak intensity of $4.0\times10^{12}$ W~cm$^{-2}$, a duration 
of $1.45$~fs (FWHM), and a sine electric-field carrier.
The central photon energy is 1~Hartree ($27.2$~eV).
The pulse has a finite duration of 4.6~fs ($t_1=70$~au[t]; $t_2=95$~au[t]; See 
eqs. 71-72 of Ref.~\cite{Patchkovskii2016}). 
We use a non-uniform grid, consisting of $10$ points with a grid spacing of 
$0.0364$~Bohr at the the origin, followed by a
$25$-point logarithmic grid with a scaling parameter of $1.1$, starting at 
$0.4$~Bohr, and a 965-point uniform grid with
a spacing of $0.4$~Bohr starting at $4.34$~Bohr. A transmission-free complex 
absorbing potential\cite{Manolopoulos02}
($k_{min}=0.2$, $\delta=0.2$) starts $357.9$~Bohr from the origin and has a 
width of $32.8$~Bohr. The energy of the initial
ground state calculated on this grid is $-0.499988$~Hartree. We use a time 
step of $0.01$~au[t]. The simulation includes
angular channels with $L\le4$, and is converged to machine accuracy with respect 
to the angular momentum and the time step.
The Coulomb potential was not modified.

\begin{figure}{}
\begin{center}
\includegraphics[width=0.48\textwidth]{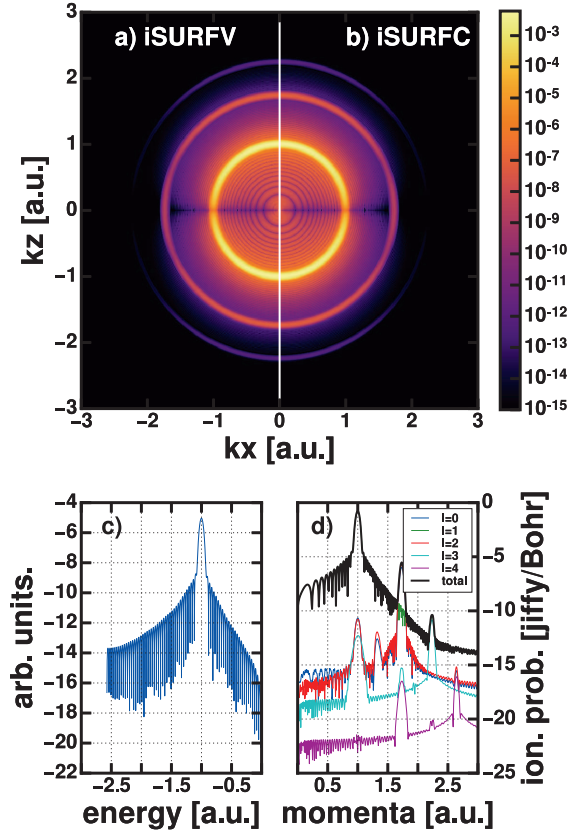}
 \caption{Angle-resolved photoelectron spectra for H(1s) and a 1.45~fs 
$4.0\times10^{12}$ W~cm$^{-2}$ pulse with a central photon energy of 27.2~eV.
          The panels are:
          (a) The left panel is for the Volkov-state projection (iSURFV);
          (b) The right panel is for the Coulomb scattering state projection 
(iSURFC);
          (c) Spectral content of the ionizing pulse;
          (d) Energy-resolved PES for the iSURFC projection.
         The background features are due to one-photon ionization by 
         the pulse wings, arising from temporal truncation of the Gaussian.
         }
 \label{fig:hfew:angular}
\end{center}
\end{figure}

\begin{figure}{}
\begin{center}
\includegraphics[width=0.48\textwidth]{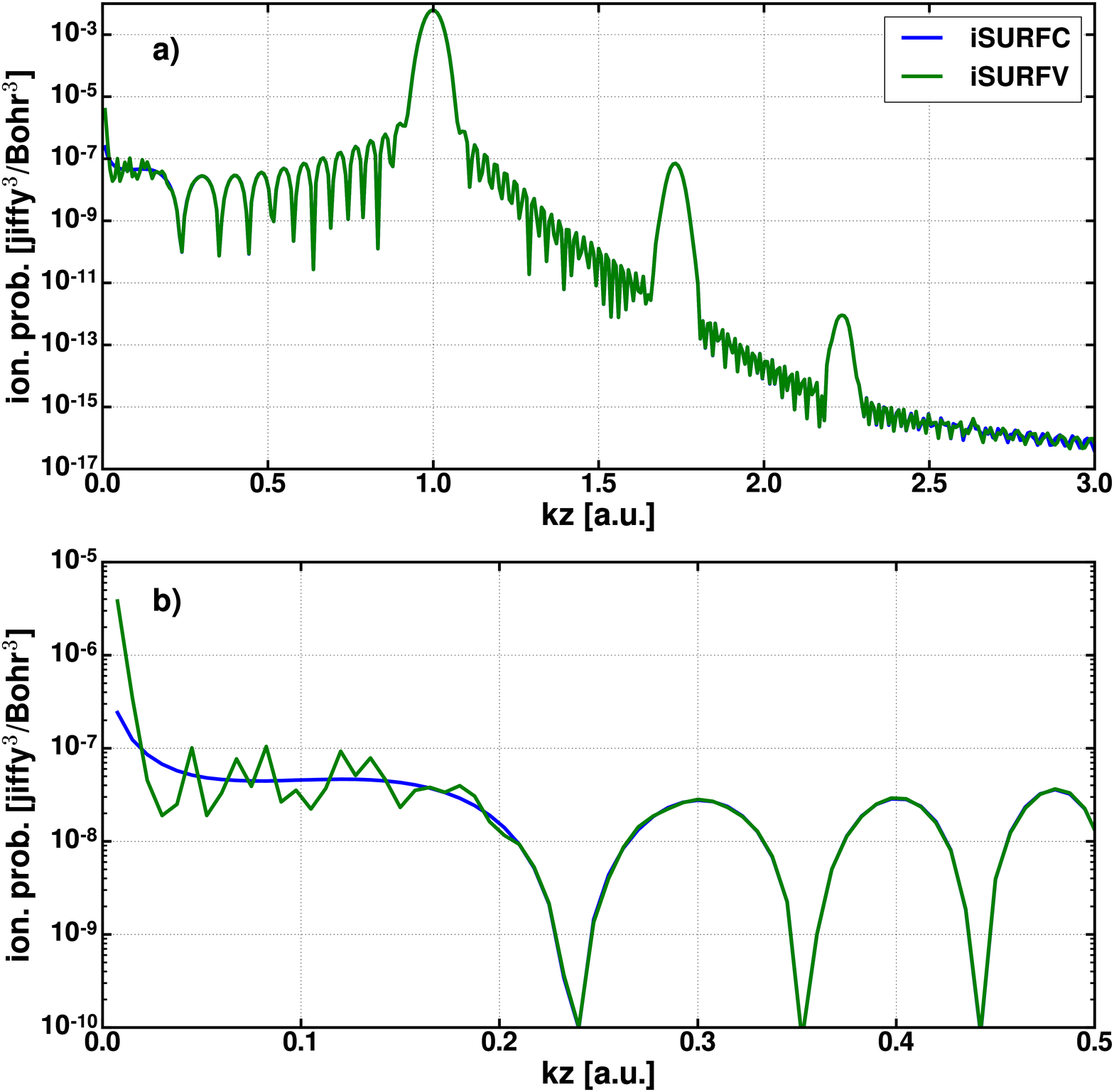}
 \caption{Cuts of the photoelectron spectra of 
Fig.~\ref{fig:hfew:angular} along the positive direction
          of the $k_z$ axis. The Coulomb-wave (iSURFC) and Volkov-state 
(iSURFV) projections are given by blue and green lines, respectively.
          Panel (a) shows the full range of photoelectron momenta;
          panel (b) shoes the low-momentum part of the spectrum.
         }
 \label{fig:hfew:section}
\end{center}
\end{figure}

The dividing surface is placed $347.9$~Bohr away from the origin, $10$~Bohr 
before the start of the absorber. 
By the end of the pulse, the entire population is still contained within the 
dividing surface. Therefore, we
only calculate the infinite-time contributions to the iSURF photoelectron 
spectra. The total probability
of ionization is $0.132$\%, with the 1-, 2-, and 3-photon ionization peaks 
clearly visible in the spectra.
The PES is cylindrically symmetric around the laser polarization direction. The 
$XZ$ section of the PES is
shown in Figure~\ref{fig:hfew:angular}. The iSURFV and iSURFC spectra are 
barely distinguishable on this scale.

The lineouts of the spectra along the positive $k_z$ direction are shown in 
Fig.~\ref{fig:hfew:section}.
For final momenta exceeding $\approx0.3$~Bohr/jiffy (kinetic energy of $\approx 
1.2$~eV) the cross-sections
(but not the phases, data not shown) calculated using the Volkov and 
Coulomb-state projection are virtually identical.
However, careful examination of the low-energy part of the spectrum shows that 
the iSURFC spectrum goes smoothly
to the expected cusp at zero energy. The Volkov projections, on the other hand, 
broadly follow the correct
cross-sections, but show an increasingly oscillatory pattern when approaching 
zero momentum. 
This defect is expected: plane waves are an increasingly poor approximation to the 
Coulomb scattering wave
for low final momenta. In principle, the range of $k$ magnitudes where 
substantial distortions occur can
be reduced by increasing the simulation volume. However, this procedure 
converges very slowly (as $O(R_{\rm max}^{-1/2})$).
Alternatively, the problem could also be alleviated by using eikonal-Volkov 
states \cite{Ermolaev00}. 
However, since eikonal-Volkov states are only approximately 
orthonormal \cite{Smirnova08}, they potentially
introduce other, hard to control, artifacts in the PES.

\subsection{Strong-field ionization of the Hydrogen atom ground 
state\label{sec:example:hstrong}}

Again, we start with the hydrogen atom in its $1s$ ground state. The structure 
of the radial grid and the parameters of
the absorbing boundary are the same as in the previous example 
(Sec.~\ref{sec:example:hfew}).
The laser field is a linearly-polarized Gaussian pulse along the Cartesian $Z$
direction, with a peak intensity of $1.0\times10^{14}$ W~cm$^{-2}$, a duration 
of $4.84$~fs (FWHM), and a sine electric-field carrier.
The central photon energy is 0.05695~Hartree (800~nm; 1.55~eV). The pulse has a finite 
duration of 11.9~fs ($t_1=170$~au[t]; $t_2=245$~au[t]~\cite{Patchkovskii2016}),
comprising approximately 5 cycles. The simulations used a time step of 
$0.0025$~jiffies and $L_{\rm max}=60$.

\begin{figure}{}
\begin{center}
\includegraphics[width=0.48\textwidth]{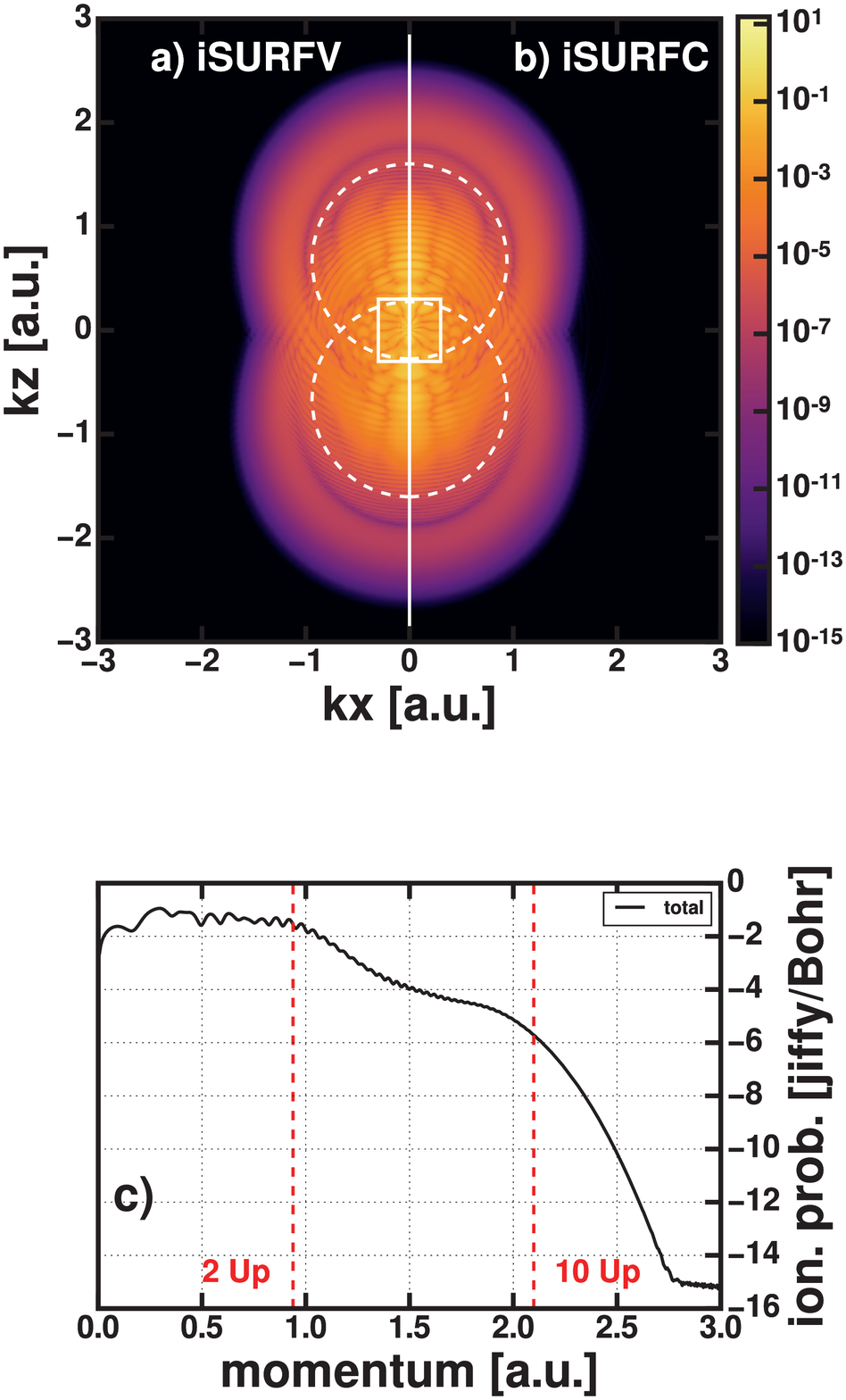}
 \caption{Angle-resolved photoelectron spectrum for H(1s) and a 4.84~fs 
$1.0\times10^{14}$ W~cm$^{-2}$, 800~nm pulse,
          with the dividing surface at $827.9$~Bohr, $10$~Bohr before the start 
of the absorber.
          The panels are:
          (a) The left panel is for the Volkov-state projection (iSURFV);
          (b) The right panel is for the Coulomb scattering state projection 
(iSURFC);
          (c) The energy-resolved iSURFC photoelectron spectrum.
         The $2 U_p$ re-scattering rings are indicated with dotted rings. 
         The iSURFV and iSURFC PES are visually indistinguishable on this 
scale. The part of the spectrum indicated by the white
         square in panel (a) is shown in Fig.~\ref{fig:hstrong:insert}.
         }
 \label{fig:hstrong:big:angular}
\end{center}
\end{figure}

\begin{figure}{}
\begin{center}
\includegraphics[width=0.48\textwidth]{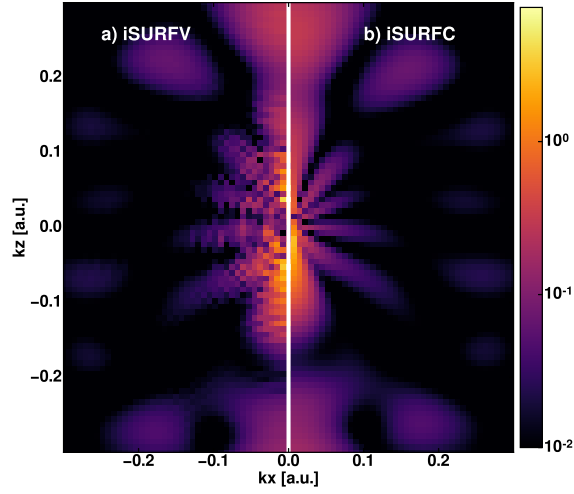}
 \caption{The low-energy structures in the photoelectron spectra for H(1s) 
strong-field ionization. The pulse parameters
          are the same as in Fig.~\ref{fig:hstrong:big:angular}.
          The panels are:
          (a) The left panel is for the Volkov-state projection (iSURFV);
          (b) The right panel is for the Coulomb scattering state projection 
(iSURFC).
         }
 \label{fig:hstrong:insert}
\end{center}
\end{figure}

\begin{figure}{}
\begin{center}
\includegraphics[width=0.8\textwidth]{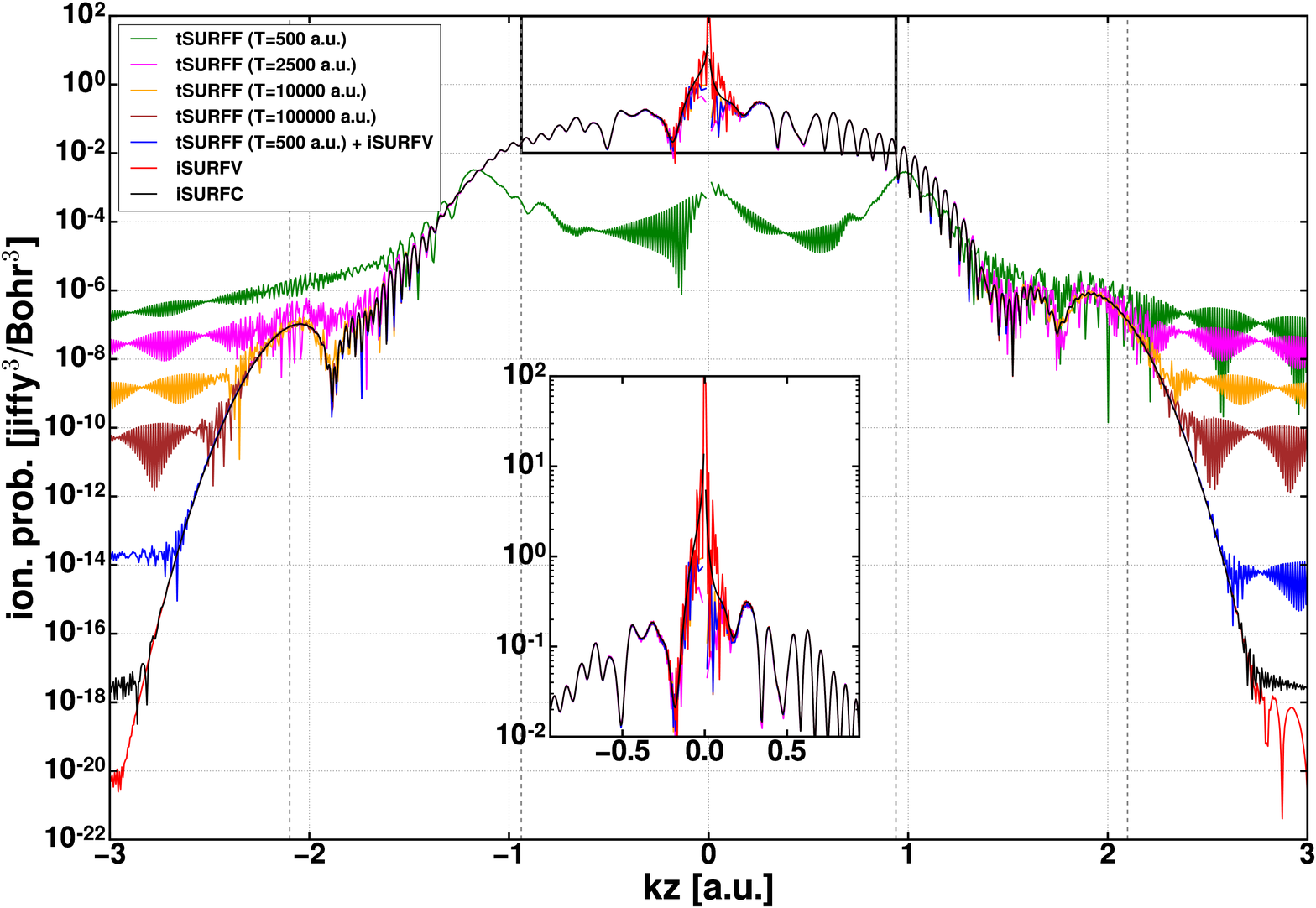}
 \caption{Photoionization probabilities for $K_x=K_y=0$, calculated with the 
tSURFF method for different time delays.
          Pulse parameters are the same as in 
Figure~\ref{fig:hstrong:big:angular}. The dividing surface is taken at
          $266$~Bohr, with the interaction potential truncated smoothly to zero 
between $240$ and $260$ Bohr.
          Photoelectron spectra sections calculated with the iSURFV and iSURFC 
(using dividing surface at $828$ Bohr),
          and the coherent sum of the tSURFF and iSURFV spectra at the end of 
the pulse are shown for comparison.
          The insert shows the PES convergence in the low-energy region for the 
iSURFC, iSURFV, and tSURFF calculations.
         }
 \label{fig:hstrong:big:cut}
\end{center}
\end{figure}

We start by calculating the PES using a large, $R_{\rm max}=870$ Bohr simulation 
box. This box is sufficient 
to contain the entire wavefunction at the end of the laser pulse. The resulting 
angle-resolved spectra are shown in
Figure~\ref{fig:hstrong:big:angular}. All the expected features \cite{becker16} of 
a strong-field PES are clearly visible,
including the multiple re-scattering rings, the holographic ``fingers'', the ATI 
rings, the low-energy structures, etc.
In the angle-integrated ATI spectrum (Figure~\ref{fig:hstrong:big:angular}c), 
the $2U_p$ and $10U_p$ cut-offs and the
low-energy structure are clearly visible. At final photoelectron momenta 
$\ge0.3$ Bohr/jiffy, the iSURFV and iSURFC
spectra are visually indistinguishable. 
Zooming into the low-energy structure region, (Figure~\ref{fig:hstrong:insert}), 
we see that all qualitative
features of the LES are present in the iSURFV spectrum. However, these features 
are superimposed into the
artefactual interferences due to the plane-wave final states, and may be 
difficult to discern without referring
to the (exact) iSURFC spectrum.

The spectra in Figures~\ref{fig:hstrong:big:angular} and 
\ref{fig:hstrong:insert} are obtained using the final-time
analysis of the total wavefunction, which remains entirely within the simulation 
box. The iSURFV spectrum is equivalent
to the infinite-time limit of the tSURFF method. It is therefore instructive to 
examine the convergence of the tSURFF
spectrum with the simulation time. The calculated cuts of the tSURFF spectra 
along the $K_x=K_y=0$ direction are
shown in Figure~\ref{fig:hstrong:big:cut}. Evaluating the tSURFF amplitudes 
immediately after the end of the laser pulse
(the green curve) does not result in a useful spectrum. After additional 
$2000$~jiffies ($\approx50$~fs; magenta curve),
the spectra are converged in the plateau region ($0.5\le K_z\le 
1.3$~Bohr/jiffy). Fully converging the spectrum within 
the second plateau region ($K_z\le2.1$~Bohr/jiffy) requires the simulation to 
continue for $10^5$~jiffies (2.5~{\it pico}seconds; brown curve).
At lower simulation times, the spectrum may appear to converge in some regions, 
but remains unconverged for similar momenta
in the opposite direction ($10^4$~Bohr/jiffy, gold curve).
Remarkably, coherently adding the iSURFV term at the end of the laser pulse to 
the tSURFF spectrum (blue line) results
in an essentially converged simulation, at the negligible additional cost 
compared to the tSURFF simulation alone (green line). We note that an ad-hoc
technique for accelerating convergence of the tSURFF time integral has also 
been proposed in \cite{Zielinski16}.

None of the tSURFF and/or iSURFV simulations converge to the correct result in 
the low-energy region (Figure~\ref{fig:hstrong:big:cut} insert).
The level of artifacts in the LES region decreases with an increasing radius of 
the matching sphere; however, the convergence
is extremely slow.

\subsection{Strong-field ionization of the ``Argon'' $3p_{-1}$ state in a circularly 
polarized field\label{sec:example:arcirc}}

Our final example involves calculation of the photoelectron spectrum of the 
$3p_{-1}$ state of the ``Argon 2P'' effective potential,
fitted to reproduce valence and Rydberg one-particle states of the Argon atom:
\begin{equation}
  v_{Ar2P}\left(r\right) = -\frac{1}{r}\left(1 + 7.625195 \ee^{-1.02557 r} - 
\frac{124.55}{1+\ee^{10\left(r-0.37110\right)}}\right).
  \label{eqn:ar2P}
\end{equation}
The laser field is a Gaussian pulse, circularly-polarized within the Cartesian 
$XY$ plane.
The pulse has a peak intensity of $2.0\times10^{14}$ W~cm$^{-2}$, a duration of 
$0.73$~fs (FWHM), and sine (X axis) and cosine (Y axis)
carrier--envelope phases. The central photon energy is 0.15~Hartree ($4.08$~eV; 
304~nm).
The pulse has a finite duration of 4.6~fs ($t_1=70$~au[t]; $t_2=95$~au[t]; See 
eqs. 71-72 of Ref.~\cite{Patchkovskii2016}). 
We use a non-uniform grid, consisting of $50$ points with a grid spacing of 
$0.0392$~Bohr at the the origin, followed by a
$104$-point logarithmic grid with the scaling parameter of $1.02$, starting at 
$2.0$~Bohr, and a uniform grid with
a spacing of $0.3$~Bohr starting at $15.4$~Bohr. A transmission-free complex 
absorbing potential \cite{Manolopoulos02} 
($k_{min}=0.2$, $\delta=0.2$) with a width of $32.8$~Bohr is used. The energy of the 
initial
ground state calculated on this grid is $-0.569$~Hartree. We use a time step 
of $0.005$~jiffy. The simulation includes
angular channels with $L\le18$, and is converged to machine accuracy with 
respect to the angular momentum and the time step.

\begin{figure}{}
\begin{center}
\includegraphics[width=0.48\textwidth]{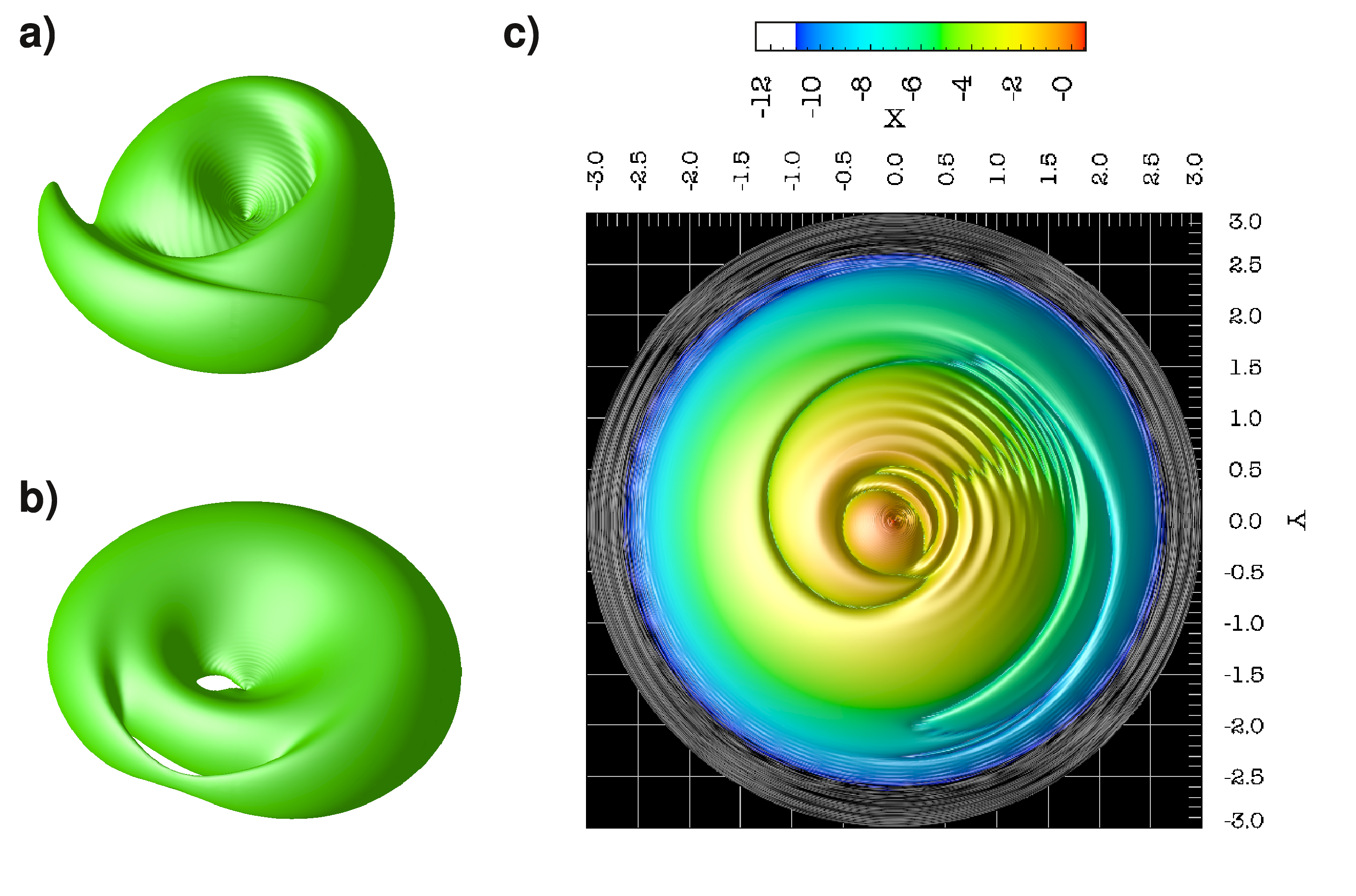}
 \caption{Magnitudes of the ionization amplitudes for the ``Argon 2P'' $3p_{-1}$ 
in a 0.73~fs, circularly polarized (XY plane) field with a peak
          intensity of $2\times10^{14}$~W~cm$^{-2}$ and a central photon energy of 
$4.08$~eV.
          a) Amplitude isosurface for the counter-rotating field;
          b) Amplitude isosurface for the co-rotating field.  The 
isosurfaces are at the $0.04$~(jiffy/Bohr)$^{-3/2}$ level;
          c) $K_x, K_y$ section of the the (a)-(b) difference. The color scheme 
is logarithmic.
         The radial box extent is $455$~Bohr. The matching sphere is at 
$413$~Bohr.
         }
 \label{fig:arcirc:3d}
\end{center}
\end{figure}

The iSURFC photoelectron spectra in the large, $455$ Bohr simulation box are 
shown in Figure~\ref{fig:arcirc:3d}. For the laser field
co-rotating with the initial state, absorption of the IR photons leads to an 
increase in the photoelectron angular momentum, following
the usual selection rules. This results in the pronounced ``doughnut'' structure 
in the 3D spectrum (panel b). The counter-rotating field,
on the other hand (panel a) can populate states with $L=0$, filling out the 
spectrum at low final momenta. In both cases, the peak of
the ionization probability is found at $\approx 0.54$~Bohr/jiffy, close to where 
the first ATI peak would be expected in a CW field
of the same intensity and frequency ($\approx 0.4$~Bohr/jiffy). Subtraction of 
the counter- and co-rotating PES (panel c) reveals
a clear ATI progression on one of the sides of the distribution, where the two 
sub-cycle bursts found in the counter-rotating case
interfere. In the opposite direction, a smooth, featureless spectrum is seen, 
with a strong energy dependence in the final emission
direction for the two fields.

\begin{figure}{}
\begin{center}
\includegraphics[width=0.8\textwidth]{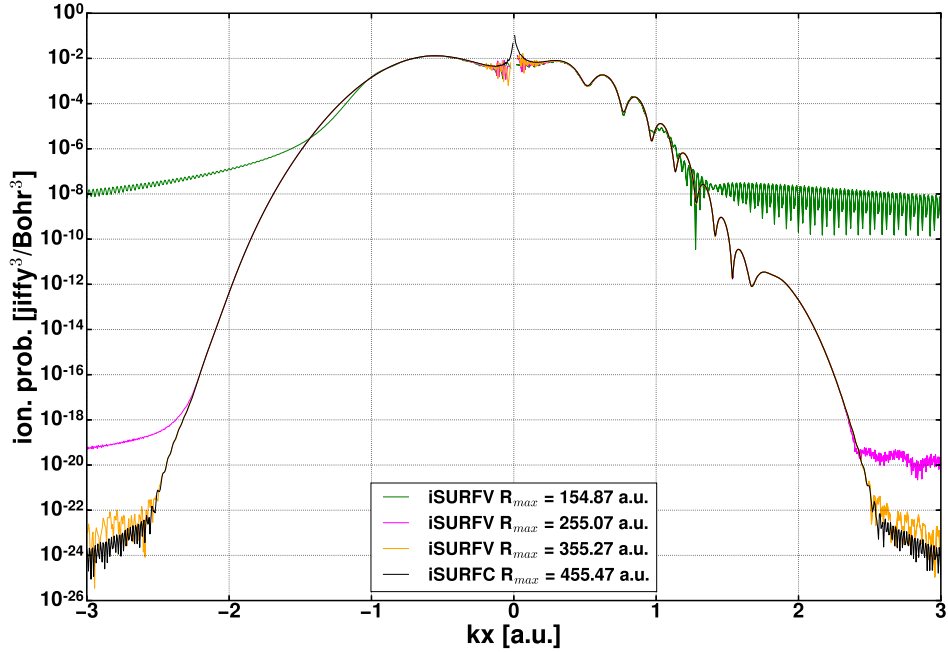}
 \caption{Cut through $K_y=K_z=0$ for the co-rotating angular distribution in 
Figure~\ref{fig:arcirc:3d}b. 
          The projection of the Coulomb states (iSURFC, black) is for the 
$455$~Bohr box. 
          The Volkov-state projections are calculated as a coherent sum of the 
``tSURFF'' projection up to the
          end of the pulse and the iSURFV infinite-time correction. The 
Volkov-state projections are for the
          boxes of $155$ (green), $255$ (magenta), and $355$~Bohr (orange). In 
all cases, the matching surface
          is placed $42$~Bohr before the end of the box.
         }
 \label{fig:arcirc:cut}
\end{center}
\end{figure}

\begin{figure}{}
\begin{center}
\includegraphics[width=0.48\textwidth]{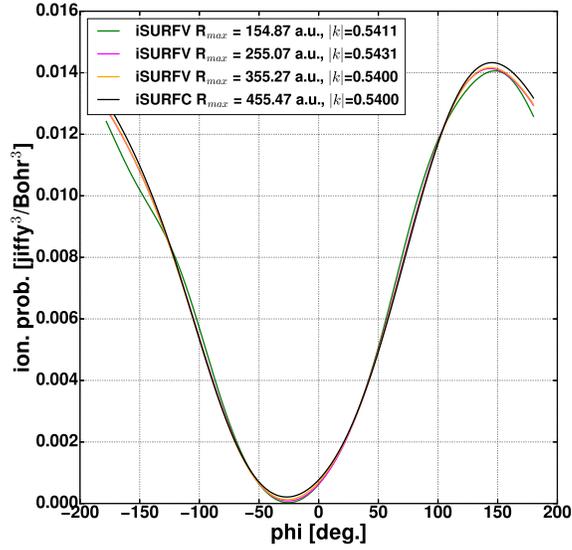}
 \caption{Cut through $K_z=0, \left|K\right|=0.54$ for the co-rotating angular 
distribution in Figure~\ref{fig:arcirc:3d}b. 
          The $K$ magnitude was adjusted for the overall shift in the potential 
due to the long-range part truncation (see text).
          The simulation parameters and line colors are the same as in 
Figure~\ref{fig:arcirc:cut}.
%           The overall dependence and the vicinity of the minimum and the maximum 
% of ionization probability
%           are shown in panels a, b, and c, respectively.
         }
 \label{fig:arcirc:ang}
\end{center}
\end{figure}

Next, we examine the convergence of the iSURFV result with respect to the 
position of the matching sphere. 
In order to guarantee that the physical Hamiltonian coincides with the Volkov 
Hamiltonian outside
of the matching sphere \cite{tao2012}, we truncate the long-range part of the 
potential by applying the transformation:
\begin{eqnarray}
  v_{\rm cut}\left(r\right) & = \left( v\left(r\right) + v_{\rm shift} \right) 
f_{\rm mask}\left(r\right) \label{eqn:mask-shift} \\
  f_{\rm mask} & = \cases{1,& $r\le R_0$\\
                          \frac{1}{2} - \frac{1}{2} f_{\rm B}\left(f_{\rm 
B}\left(f_{\rm B}\left(2\frac{r-R_0}{r-R_{\rm X}}-1\right)\right)\right),
                          & $R_0\le r\le R_{\rm X}$ \\
                          0, & $r\ge R_{\rm X}$} 
                \nonumber \\
  f_{\rm B} & = \frac{3}{2} x - \frac{1}{2} x^3. \nonumber
\end{eqnarray}
In eq.~\ref{eqn:mask-shift}, $f_{\rm B}$ is the Becke's switching 
function \cite{Becke88a}. The resulting shifted and masked potential $v_{\rm 
cut}$ is
smooth to $O\left(r^7\right)$. The masking radii $R_0$ and $R_{\rm X}$ are 
chosen $22$ and $2$~Bohr before the matching sphere. The vertical shifts 
are $9.3$, $4.89$, and $3.28$~mH, respectively for the simulation boxes of 
$155$, $255$, and $355$~Bohr. 

The cuts in the calculated photoionization probabilities along the $K_x=K_z=0$ 
direction are collected in Figure~\ref{fig:arcirc:cut}.
The Volkov-state projection calculated with the smallest matching sphere radius 
($155$~Bohr, green line) agrees well with the ``exact'' iSURFC
spectrum for final momenta between $0.3$ and $1.0$ Bohr/jiffy. Matching further 
away from the origin ($255$~Bohr, magenta line) leads to a
virtually converged spectrum for $0.2\le|K|\le2.4$ Bohr/jiffy. Finally, Volkov 
projection at $355$~Bohr (gold line) is identical to the
exact result for $|k|\ge0.2$, and is limited by the numerical accuracy of the 
underlying type.

The situation however changes if one considers angular-resolved distributions. 
In Figure~\ref{fig:arcirc:ang} we show constant-momentum
cuts of the ionization probability in the $K_z=0$ plane. Because the three 
tSURFF/iSURFV simulations use slightly different potential
shifts (see above), the cuts are taken at slightly different $|K|$ values, 
corresponding to absorption of the same energy from the laser
field. Because the photoelectron distributions in this range of $K$ are smooth, the 
exact position of the $K$ cut does not materially affect
the discussion. The maximum and minimum of the ``exact'' distribution are found 
at respectively $145.5$ and $-26.5$ degrees with respect to the 
laboratory $X$ axis direction.
For the smallest ($155$~Bohr) matching sphere, the Volkov projection leads to the 
maximum appearing at $148.0$ degrees, with the minimum at $-26.0$ degrees.
More importantly the shape of the angular dependence is substantially different. 
Increasing the size of the matching sphere moves the position of
the maximum to $144.0$ for both $255$ and $355$ Bohr spheres. The minimum is 
found at $-26.0$ degree for both spheres. 
While a $1.0$ degree deviation in the position of the maximum does not appear to 
be critical, for $1.6\mu$m driving field, for example, this would
correspond to a $15$-attosecond error in an attoclock measurement \cite{eckle08} -- 
comparable to the measured time delays \cite{pfeiffer2012}.
Furthermore, the convergence of the tSURFF and iSURFV angular distributions 
with the matching-sphere radius appears to be extremely slow.

\section{Conclusions}

We show that the surface-flux approach for calculation of photoelectron spectra 
\cite{Ermolaev99a,Ermolaev00,Serov01a,tao2012} allows natural, analytical continuation
to infinite time. For large-box simulations, where the entire wavefunction 
remains within the simulation volume at the end of the pulse,
the infinite-time form can be used to evaluate the ``exact'' ionization 
probabilities and phases in both long- and short-range potentials (the iSURFC 
approach).
The knowledge of the asymptotic form of the scattering solutions is sufficient 
for these calculations; it is not necessary to evaluate scattering
states in the vicinity of the origin. 

For small simulation volumes, where parts of the electron probability reach the 
absorber while the laser field is still on, it is no longer
possible to evaluate the projection onto the exact scattering states. However, the 
infinite-time continuation can still be applied to the Volkov states
(becoming plane waves in the absence of the laser field). This correction can be 
combined coherently with the Volkov-state continuation (``tSURFF''),
yielding the iSURFV approach. This technique produces the same, well-documented 
artifacts \cite{tao2012} as the ``tSURFF'' approach. However, it
affords the fully-converged projection onto the Volkov states immediately after 
the end of the laser pulse, without the need for tedious field-free propagation.

\section{Acknowledgments}
We would like to thank Armin Scrinzi for inspiring discussions and 
for his helpful advice. We would also like to thank Misha Ivanov for
his support and encouragement. We would like to acknowledge support from the  
Deutsche Forschungsgemeinschaft project SM 292/2-3.

\section{Appendix}
\subsection{Implementation of tSURFF and iSURFV}

In \textsc{scid-tdse}, the time-dependent wavefunction has the form:
\begin{equation}
 \Psi(\vec{r},t) = \frac{1}{r}  \sum_{LM} \Psi^R_{LM} (r,t) Y_{LM} (\Omega_r).
\label{eqn:oneparticle-wf}
\end{equation}
The Volkov functions are given by:
\begin{equation}
 X=\frac{1}{(2 \pi)^{2/3}} exp\left(i\vec{k}\vec{r}\right) exp\left(-\frac{i}{2\hbar m} \int_{t_i}^{t} (\hbar \vec{k} - e \vec{A})^2 dt\right),
\label{eqn:volkovfunc1}
\end{equation}
where the plane wave term can be expressed in a spherical harmonics expansion (see \cite{varshalovich1988}, section 5.17, formula 14, page 165)
\begin{equation}
 exp\left(i\vec{k}\cdot\vec{r}\right) = 4 \pi \sum_{LM} i^L j_L(kr) Y_{LM}(\Omega_k) Y^*_{LM}(\Omega_r),
\label{eqn:planewave-spherical}
\end{equation}
where $j_L(kr)$ are the spherical Bessel functions.

Then, the integral in eq.\ref{eqn:projection:surface} becomes:
\begin{equation}
\int dr X^* \frac{i}{\hbar} [ \hat{H}_S,\Theta] \Psi \nonumber = \sum_{LM}  \{ F_{LM} \Psi^R_{LM} + G_{LM} \frac{\partial}{\partial r} \Psi^R_{LM} \},
\label{eqn:tsurf-fgc}
\end{equation}
where
\begin{eqnarray}
F_{LM} = \frac{c^*r}{m} \left(-i\right)^{L+1} \left[ \frac{\hbar k }{2} Y_{LM} \left(\Omega_k\right) \frac{1}{2L+1} \left(L j_{L+1}\left(kr\right) - \left(L+1\right) j_{L-1}\left(kr\right) \right)  \right. \nonumber \\ 
\left. + eA_z \left(C_{L,M} Y_{L-1,M}\left(\Omega_k\right)j_{L-1}\left(kr\right) - C_{L+1,M} Y_{L+1,M}\left(\Omega_k\right)j_{L+1}\left(kr\right) \right) \right] \label{eqn:f-function-fgc},  
\end{eqnarray}
\begin{equation}
G_{LM} = \frac{c^*r}{m} \left(-i\right)^{L+1} \frac{\hbar }{2} Y_{LM} \left(\Omega_k\right) j_{L}(kr) \label{eqn:g-function-fgc}, 
\end{equation}
\begin{equation}
C_{LM} = \left( \frac{L^2 - M^2}{4L^2 -1} \right)^{1/2} \label{eqn:clm-function-fgc},
\end{equation}
and
\begin{equation}
c^{*} = \frac{4 \pi}{\left(2 \pi\right)^{3/2}}  exp \left(\frac{i}{2\hbar m} \int^t_{t_i} \left(h\vec{k} -e\vec{A}\right)^2 dt \right)  \label{eqn:cprime-function-fgc}. \\
\end{equation}

\subsection{Implementation of iSURFC}
For the Coulomb-state projection in eqs. \ref{eqn:eigenstate:factor},\ref{eqn:projection:infinity}, it is convenient to work
with the outgoing Coulomb spherical waves. A numerically accurate implementation
of the Coulomb spherical waves and their derivatives is available in \cite{barnett1982}.
The surface integral in eq. \ref{eqn:eigenstate:factor} can then be evaluated directly as written.

For comparison to experimental angle- and energy-resolved spectra it is then
necessary to project the outgoing spherical Coulomb wavepacket onto a Rutherford 
scattering state \cite{flugge94}. The Coulomb wavepacket with the radial wavevector $k$
is given by:
\begin{equation}
 \Psi_k(\vec{r}) = \frac{1}{r} \sum_{LM} a_{kLM} Y_{LM} (\Omega_r) F_{kLM} (r),
\end{equation}
where $F_{kLM} (r)$ are Coulomb spherical waves.
$F_{kLM}$ are normalized to $\delta \left( \frac{k-k'}{2 \pi} \right)$.
At large r, it is given by \cite{flugge94}:
\begin{equation}
 F_{kLM} \rightarrow 2 sin\left(kr + \frac{z}{k}log\left(2kr\right) - \frac{\pi L} {2} + \varphi_{L} \right),
\end{equation}
where
\begin{equation}
\varphi_{L} = Arg \left(\Gamma \left(L+1 - \frac{i z}{k}\right)\right).
\end{equation}

The Rutherford scattering solution (the ``Coulomb plane wave'') normalized $\delta(\vec{k}-\vec{k'})$, 
is asymptotically given by \cite{flugge94}:
\begin{eqnarray}
 w\left(\vec{k}\right) = \sqrt{\frac{2}{\pi}} \frac{1}{k'r} \sum_{L'M'} i^{L'} exp \left( i\varphi_{L'} \right) sin\left(k'r + \frac{z}{k'}log\left(2k'r\right) - \frac{\pi L} {2} + \varphi_{L'} \right) \nonumber \\ 
Y^*_{LM} \left(\Omega_k\right) Y_{LM} \left( \Omega_r \right)
\end{eqnarray}

Calculating an overlap between $\Psi_k$ and $w(\vec{k})$, we immediately obtain amplitudes $b_{\vec{k}}$ 
of the Rutherford states:
\begin{equation}
 b_{\vec{k}} = \sum_{LM} \sqrt{\frac{1}{2\pi}} \frac{1}{k} (-i)^L exp \left(-i\varphi_L\right) Y_{LM}\left( \Omega_k \right) a_{LM}.
\end{equation}

\subsection{Evaluation of spherical harmonics}

The calculation of photoelectron spectrum requires the repeated evaluation of spherical harmonics
$Y_{LM}$, potentially for high values of angular momenta $L$ and $M$. We find that the following
recurrence formulas are fast and numerically stable for large $L$ and $M$:
\begin{equation}
 \label{y00}
Y_{00} = \frac{1}{\sqrt{4 \pi}},
\end{equation}
\begin{eqnarray}
 \label{ylmone_lmone}\nonumber \\
\frac{Y_{L,L}}{Y_{L-1,L-1}} = i \sqrt{\frac{2L+1}{2L}} exp \left(i \varphi \right) sin \left(\theta\right), \nonumber \\
\frac{Y_{L,-L}}{Y_{L-1,-(L-1)}} = -i \sqrt{\frac{2L+1}{2L}} exp \left(-i \varphi \right) sin \left(\theta\right),
\end{eqnarray}
and finally:
\begin{equation}
 \label{yLM}
Y_{LM} = iz \sqrt{\frac{4L^2-1}{L^2 - M^2}} Y_{L-1,M} + \sqrt{\frac{2L+1}{2L-3} \frac{\left(L-1\right)^2 - M^2}{L^2-M^2}} Y_{L-2,M}.
\end{equation}
These recurrences formulas allow for the calculation of a range of $L$ values for a fixed $M$, 
without having to evaluate all intermediate $M$.
\newpage
\bibliography{scid-tdse-isurf}

\end{document}